\documentclass[preprint,amsmath,amssymb, aps, prl,
floatfix,nofootinbib]{revtex4-2} 
\usepackage{graphicx}
\usepackage{dcolumn}
\usepackage{bm}
\usepackage{caption}
\usepackage{amssymb,amsmath,color}
\usepackage{gensymb}
\usepackage{url}
\usepackage{float}
\usepackage{braket}
\usepackage{hyperref}    
\hypersetup{colorlinks = true, linkcolor = blue, anchorcolor = blue, citecolor = blue, filecolor = blue, urlcolor = blue}
\usepackage{xcolor}
\usepackage{float}
\usepackage{tikz}
\usepackage{pgfplots}

\begin{document}

\title{Chiral valley phonons and flat phonon bands in moir\'{e}
materials}

\author{Indrajit Maity, Arash A. Mostofi} \author{ Johannes
Lischner} \email{j.lischner@imperial.ac.uk}

\affiliation{Departments of Materials and Physics and the Thomas
Young Centre for Theory and Simulation of Materials, Imperial
College London, South Kensington Campus, London SW7 2AZ, UK}%

\begin{abstract}
We investigate the chirality of phonon modes in twisted bilayer
$\mathrm{WSe_{2}}$ and demonstrate distinct chiral behavior of the
$K/K^\prime$ valley phonons for twist angles close to
$0^{\circ}$ and close to $60^{\circ}$. In particular, multiple chiral non-degenerate $K/K^\prime$ valley phonons are found for twist angles near $60^{\circ}$ whereas no non-degenerate chiral modes are found for twist angles close to $0^\circ$. Moreover, we discover two sets of emergent chiral valley modes that originate from an inversion symmetry breaking at the moir\'{e} scale and find similar modes in moir\'{e} patterns of
strain-engineered bilayers $\mathrm{WSe_{2}}$ and
$\mathrm{MoSe_{2}/WSe_{2}}$ heterostructures. At the energy gap between acoustic and optical modes, the formation of flat phonon bands for a broad range of twist angles is observed in twisted bilayer
$\mathrm{WSe_{2}}$. Our findings are relevant for understanding
electron-phonon and exciton-phonon scattering in moir\'e materials and also for the design of phononic analogues of flat band electrons.

\end{abstract}

\maketitle 

\textit{Introduction.} Chirality - the characteristic of an object that can be
distinguished from its mirror image - plays a fundamental role in physics, chemistry, and biology. For example, different enantiomers of chiral molecules absorb different amounts of right-handed and left-handed polarized light, which enables their spectroscopic characterization through measurement of the circular dichroism~\cite{Kneercircular2018}. In condensed matter physics, the chirality of electrons gives rise to many exotic effects, such as Klein tunneling~\cite{Katsnelsonchiral2006} in graphene, and the chiral magnetic effect in three-dimensional semi-metals~\cite{Chiralqiang2016,Qiangchiral2016}.

Recently, the study of chirality in phonons has attracted significant interest. For example, Zhang and Niu~\cite{Zhangchiral2015} predicted the existence of chiral phonons at the $K$ and $K^\prime$ valleys of monolayer $\mathrm{WSe_{2}}$, which was subsequently verified by experiments that measured the circular dichroism of phonon-assisted intervalley transitions of holes~\cite{Zhuobservation2018}. The interaction of such chiral valley phonons with other quasiparticles is relevant for understanding and controlling many electronic and optical phenomena~\cite{Carvalhointervalley2017,Chenhelicity2015,Kangholstein2018,Liemerging2019,Liuvalley2019,Limomentum2019,Delhommeelipping2020}. For instance, Li \textit{et al.} demonstrated that the coupling between a chiral valley phonon and an intervalley exciton in monolayer $\mathrm{WSe_{2}}$ can lead to a long exciton lifetime maintaining the valley polarization, which is important for valley-excitonics~\cite{Limomentum2019}. Chen \textit{et al.} observed the entanglement of  chiral phonon modes in monolayer $\mathrm{WSe_{2}}$ and single photons emitted from an embedded quantum dot in the material, which promises to reveal new avenues for phonon-driven entanglement of quantum dots~\cite{Chenentanglement2019}.

Besides monolayer $\mathrm{WSe_{2}}$, chiral phonons have been found in other materials~\cite{Chenchiral2018,Dominikorbital2019,Luojunchirality2019,Zhangchiral2020, Grissonnanchechiral2020}. A particularly promising platform for observing and manipulating chiral phonons are twisted bilayers of two-dimensional (2D) materials. Since the discovery of flat electronic bands in twisted bilayer graphene~\cite{Bistritzermoire2011}, such moir\'{e} materials have emerged as rich systems to investigate the properties of correlated electrons, excitons and phonons~ \cite{Caounconventional2018,Caocorrelated2018,Tranmoire2020,Wangcorrelated2020,Huanggiant2020,Scurielectrically2020,Naikultraflat2018,Zhangflat2020,Aninteraction2020,Xianrealization2020,Devakulmagic2021,Vitaleflat2021,Somepallielectronic2020,Lilattice2021,Sorianospin2021,Shabanideep2021,Bremhybridized2020,Westonatomic2020,Sinhabulk2020,Goodwinflat2021,Abigailghost2020,Pizarrothe2019}. For example, at small twist angles phonon properties are significantly modified as a consequence of large atomic reconstructions~\cite{Maityreconstruction2021,Westonatomic2020,Carrrelaxation2018} which result in the localization of incipient phason modes \cite{Maityphonons2020} and optical phonon modes~\cite{Maityphonons2020,Gadelhalocalization2021,Koshinomoire2019,Quanphonon2021,Lamparskisoliton2020,Linlarge2021,Debnathevolution2020}. To date, however, the chirality of phonons in twisted bilayer systems has not been studied.

In this Letter, we study the chirality of phonon modes in moir\'{e} materials. Focusing on twisted bilayer $\mathrm{WSe_{2}}$, we first demonstrate qualitatively different phonon chiralities for twist angles close to $0^\circ$ and
close to $60^\circ$ and explain these differences by analyzing the folding of the monolayer $K$ and $K^\prime$-valleys into the Brillouin zone of the twisted bilayer. For twist angles near $60^\circ$, we find two sets of emergent chiral phonons modes which originate from symmetry breaking at the moir\'e scale. Finally, we discover flat chiral phonon bands in the energy gap between acoustic and optical modes without the requirement of a magic angle. Such flat phonon bands have recently been reported in metamaterials~\cite{Dengmagic2020, Rosendoflat2020, Minhalsimulating2021, Martdipolar2021}, but not yet in moir\'e systems.

\textit{Methods.} For twist angles near $0\degree$ ($\theta$) or $60\degree$ ($60^\circ - \theta$), the size of the moir\'e unit cell becomes very large making standard first-principles approaches for calculating phonon properties unfeasible. We therefore use simpler models for interatomic interactions. 
Specifically, intralayer interactions in $\mathrm{WSe_{2}}$ and $\mathrm{MoSe_{2}}$ are described using a Stillinger-Weber potential, which has been demonstrated to give accurate results for monolayers~\cite{Arashvalidation2018,Kandemirthermal2016}. For interlayer interactions, we use the Kolmogorov-Crespi potential that can correctly reproduce the interlayer binding energy landscape obtained using first-principles calculations~\cite{Naikkolmogorov2019}. The relaxed atomic positions of our twisted bilayer systems are determined using the implementation of these potentials within the LAMMPS code~\cite{Plimptonfast1995}. Phonon frequencies and polarization vectors are obtained by diagonalizing the dynamical matrix using a modified version of the PHONOPY code~\cite{Togofirst2015} (see Supplementary Information (SI), Sec. I, for additional details). For monolayer graphene, we use the REBO potential~\cite{Brennersecond2002}.

Following Zhang and Niu~\cite{Zhangchiral2015}, we define the chirality of a phonon with crystal momentum $\mathbf{q}$ and band index $n$ along the $z$-direction (assuming that the 2D material lies in the $x$-$y$ plane) as 
\begin{equation}
\begin{split}
S^{z}_{n{\bf q}} = & \bra{\epsilon_{n{\bf q}}} \widehat{S}^{z} \ket{\epsilon_{n{\bf q}}} \\
= & \bra{\epsilon_{n{\bf q}}} \sum_{j=1}^{N} \Big[ \ket{R_{j}}\bra{R_{j}} - \ket{L_{j}}\bra{L_{j}} \Big] 
\ket{\epsilon_{n{\bf q}}},
\end{split}
\label{eq1}
\end{equation}
where $\ket{\epsilon_{n{\bf q}}}$ is the normalized polarization vector, $\widehat{S}^{z}$ denotes the phonon circular polarization operator and $N$ is the number of atoms in the moir\'e unit cell. $\ket{R_{j}}$ and $\ket{L_{j}}$ denote the right and
left circularly polarized basis vectors for atom $j$. For
example, for the first two atoms these basis vectors are given by $|R_{1} \rangle =1/\sqrt{2} \times {(1, i, 0,
0,\ldots)^{\mathrm{T}}}$, $|L_{1} \rangle = 1/\sqrt{2} \times {(1, -i, 0, 0, \ldots)^{\mathrm{T}}}$, $|R_{2} \rangle =1/\sqrt{2} \times {(0, 0, 1, i, 0, 0, \ldots)^{\mathrm{T}}}$, and $|L_{2}
\rangle = 1/\sqrt{2} \times {(0, 0, 1, -i, 0, 0, \ldots)^{\mathrm{T}}}$. 
The phonon mode is linearly
polarized when $S_{n\bf{q}}^{z}=0$, circularly polarized when $S_{n\bf{q}}^{z}=\pm 1$ and elliptically polarized when $0<|S_{n\bf{q}}^{z}|<1$. We refer to both the circularly and
elliptically polarized modes as chiral phonons. Note that for systems with time-reversal symmetry, we have $S^{z}_{n{\bf q}}=-S^{z}_{n-{\bf q}}$, whereas systems with
inversion-symmetry obey $S^{z}_{n{\bf q}}=S^{z}_{n-{\bf q}}$. Therefore, the existence of \textit{non-degenerate} chiral phonon modes requires the breaking of one of these symmetries~\cite{Zhangchiral2015,Zhuobservation2018}.

As a test, we first study the chirality of phonons in monolayer graphene (which has both time-reversal and inversion symmetry) and monolayer
$\mathrm{WSe_{2}}$ (which is not inversion symmetric). The results are shown in Sec. II of the SI~\cite{SI}. As expected, all graphene phonons are achiral. For $\mathrm{WSe_{2}}$, we find multiple \textit{non-degenerate} chiral phonon modes near the $K$ and $K^{\prime}$
valleys, consistent with previous first-principles calculations~\cite{Zhuobservation2018}, and the signs of the chiralities in the two valleys are opposite.

\begin{figure}[ht!]
\centering
\includegraphics[width=0.5\textwidth]{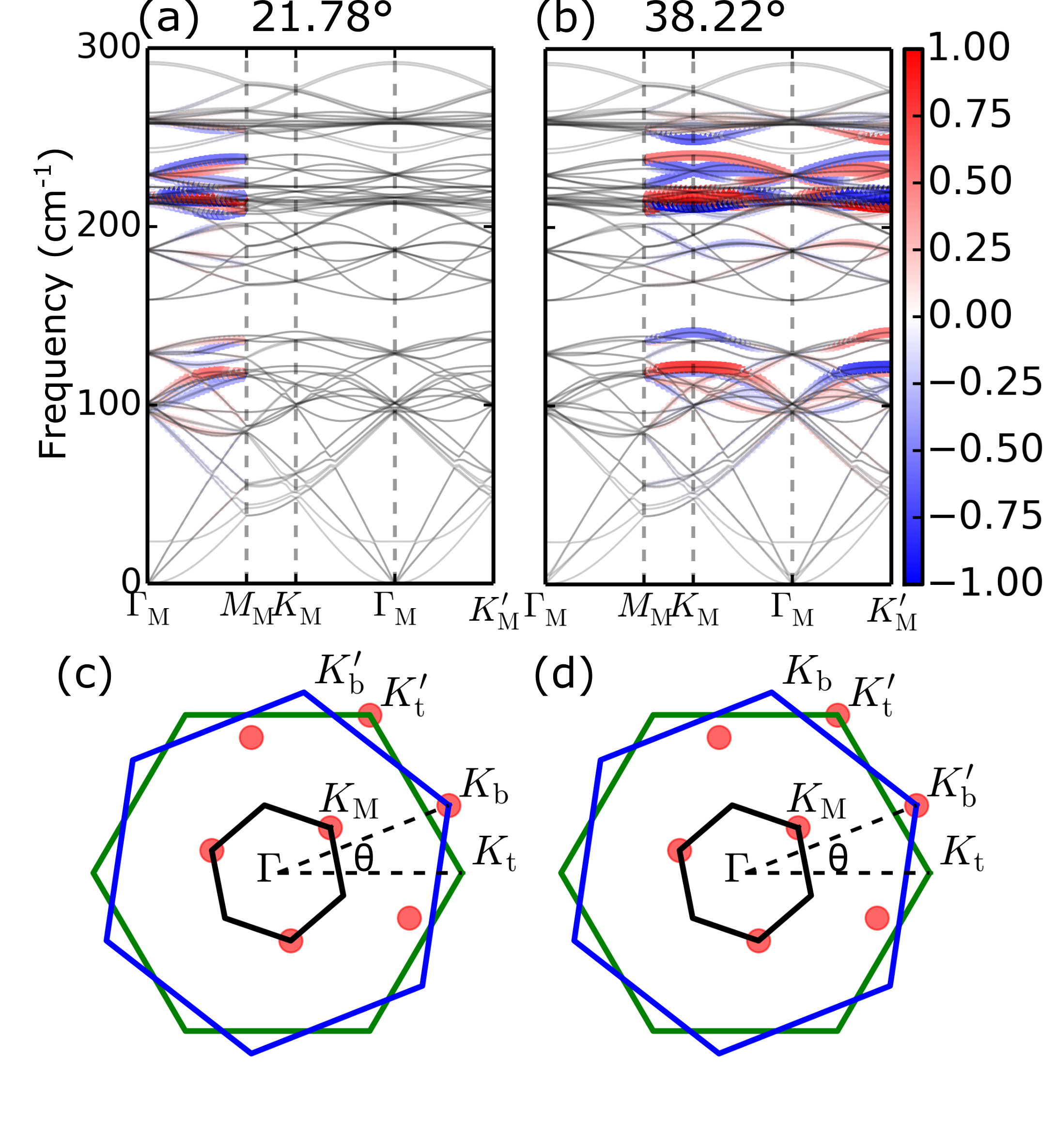}
\caption{(a),(b): Chiral phonons in twisted bilayer $\mathrm{WSe_{2}}$ at twist angles of
$21.78^\circ$ and $38.22^\circ$, respectively. Dark solid lines represent phonon bands and blue and red color indicates the associated chirality.
(c),(d): Red circles
denote the unfolded $K_{\mathrm{M}}$ point of the moir\'{e} BZ into the
monolayer BZ for $21.78^{\circ}$ and $38.22^\circ$, respectively.
The green and blue solid lines represent the unrotated and rotated monolayer
BZs, whereas the black solid line represents the moir\'{e} BZ. }  \label{fig1}
\end{figure}

\textit{Chiral valley phonons in twisted bilayer ${WSe_{2}}$.} The naturally occuring bilayer $\mathrm{WSe_{2}}$ has 2H stacking and is inversion symmetric \cite{Jonesspin2014}. The untwisted bilayer, therefore, does not exhibit chiral valley phonons. However, the introduction of a twist between the layers results in the emergence of chiral phonons, as demonstrated in
Figs.~\ref{fig1}(a) and (b), which show the phonon chirality $S^z$ of each phonon mode at twist angles of 
$21.78^\circ$ and $38.22^\circ$, respectively. At $21.78^\circ$, all phonons near the $K_{\mathrm{M}}$ and $K^\prime_{\mathrm{M}}$ valleys are achiral (with the subscript $\mathrm{M}$ denoting that these k-points belong to the moir\'e Brillouin zone), but there are some chiral phonons between $\Gamma_{\mathrm{M}}$ and $M_{\mathrm{M}}$. In contrast, we find several chiral phonon modes in the $K_{\mathrm{M}}$/$K^\prime_{\mathrm{M}}$ valleys for the $38.22^\circ$ twist angle (but none between $\Gamma_{\mathrm{M}}$ and $M_{\mathrm{M}}$). Qualitatively similar results are found for other twist angles near $0\degree$ and $60\degree$, respectively.

The different behaviour of the $K_{\mathrm{M}}$/$K^\prime_{\mathrm{M}}$ valley phonons for twist angles near $0^\circ$ and near $60^\circ$ can be understood by analyzing the folding of the monolayer phonon modes into the smaller moir\'e Brillouin zone. For the untwisted system (i.e., $\theta=0^\circ$), the $K$ point of the
top layer (denoted $K_{\mathrm{t}}$) coincides with the $K$ point of the bottom layer (denoted $K_{\mathrm{b}}$) and similarly $K^{\prime}_{\mathrm{t}}$ and $K^{\prime}_{\mathrm{b}}$ coincide. When instead a small, but finite twist angle is considered, the BZs of the two layers are rotated relative to each other and a smaller moir\'e BZ is created, see Fig.~\ref{fig1}(c). To understand which crystal momenta $\mathbf{q}_i$ of the monolayer BZs fold onto a specific point $\mathbf{Q}_{\mathrm{M}}$ in the moir\'e BZ, we ``unfold" $\mathbf{Q}_{\mathrm{M}}$ by adding all possible reciprocal lattice vectors ${\bf G}_i$ corresponding to the moir\'e crystal according to~\cite{Popescuextracting2012}
\begin{equation}
{\bf q}_{i} = {\bf Q}_{\mathrm{M}} + {\bf G}_{i}.
\end{equation} 
Figure~\ref{fig1}(c) shows the set of points in the monolayer BZs that result from unfolding $K_{\mathrm{M}}$ for a twist angle of $21.78^\circ$. It can be seen that both $K^{\prime}_{\mathrm{t}}$ and $K_{\mathrm{b}}$ fold onto $K_{\mathrm{M}}$. Since the chirality of the phonons at $K^{\prime}_{\mathrm{t}}$ and $K_{\mathrm{b}}$ have equal magnitudes, but opposite signs, the phonons of the twisted system have a vanishing chirality.

Instead, for the system with $60^\circ$ twist angle, $K_{\mathrm{t}}$ coincides with $K^{\prime}_{\mathrm{b}}$ and similarly $K^{\prime}_{\mathrm{t}}$ and $K_{\mathrm{b}}$ coincide. Fig.~\ref{fig1}(d) shows that for a twist angle near $60^{\circ}$, $K_{\mathrm{M}}$ unfolds onto $K^\prime_{\mathrm{t}}$ and
$K^{\prime}_{\mathrm{b}}$ which have phonons with the same chirality sign. As a consequence, the phonons of the twisted bilayer exhibit a non-vanishing chirality at the $K_{\mathrm{M}}$ valley (and also for the $K^\prime_{\mathrm{M}}$ valley).

\begin{figure}[!ht]
\centering
\includegraphics[width=0.5\textwidth]{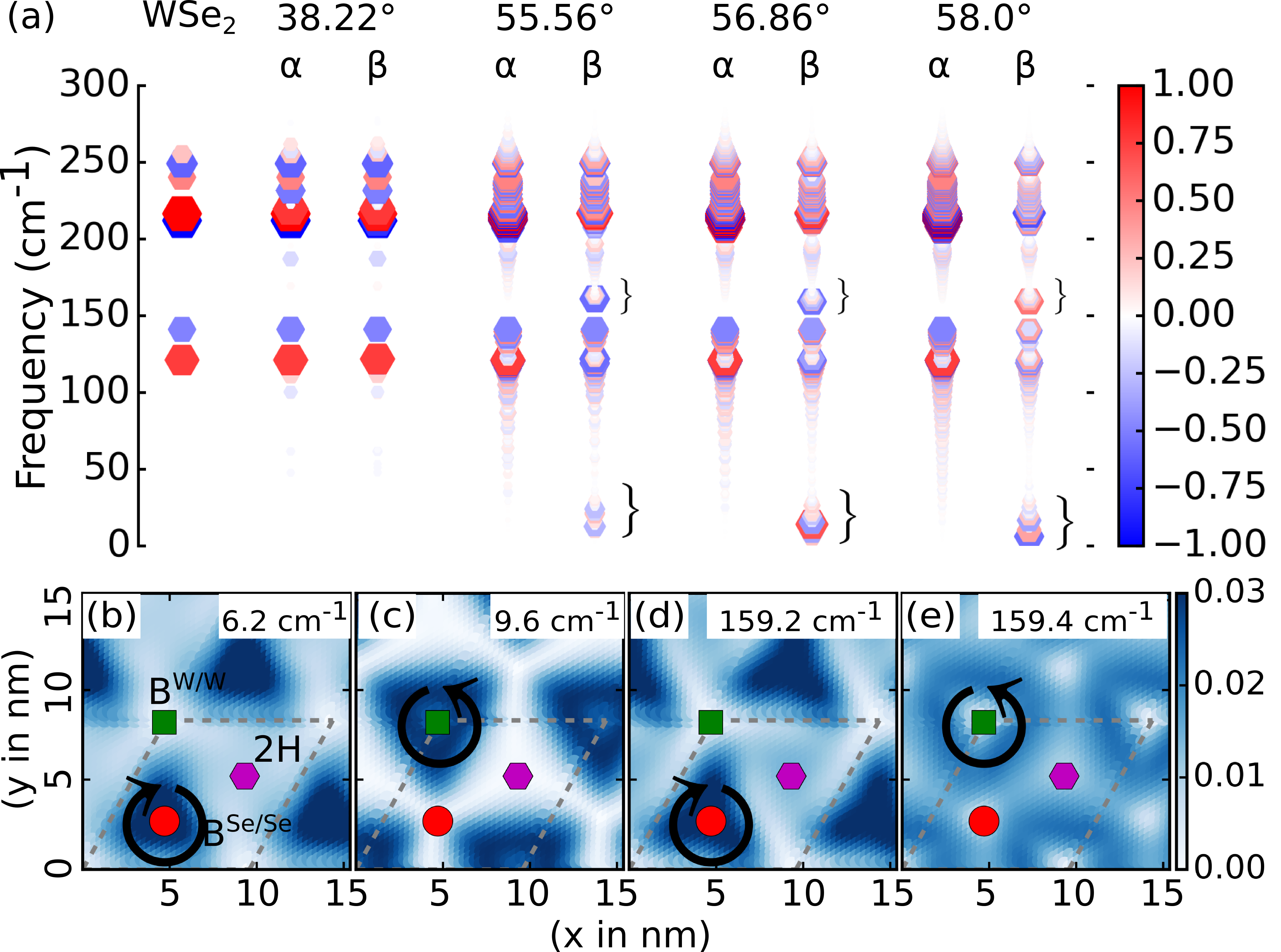}
\caption{(a) Chiral $K$-valley phonon modes of monolayer $\mathrm{WSe_{2}}$ (first column) and twisted bilayer $\mathrm{WSe_{2}}$ at different twist angles (columns labelled $\beta$). For comparison, results are also shown for twisted bilayers without interlayer interactions (columns labelled $\alpha$). The colour of the symbols indicates the chirality of the phonon modes. The emergent chiral modes are highlighted by curly brackets.
(b)-(e) Absolute value of the in-plane components of the normalized polarization vector
for some of the chiral modes for a twist angle of $58^\circ$. In (b) and (c), the polarization vector of W atoms in the
top layer is used, In (d) and (e), the polarization vector of Se atoms in the top layer is shown. Different
stacking locations and associated chiralities within the moir\'{e} unit cell,
which is denoted by dashed lines, are indicated.}  \label{fig2}
\end{figure}

Figure~\ref{fig2}(a) shows the chiralities of phonons in the $K_{\mathrm{M}}$-valley as a function of twist angle near $60\degree$ (columns labelled $\beta$) and compares them to the $\mathrm{WSe_{2}}$ monolayer result (the first column) and also to the results of \emph{decoupled} twisted bilayers at the same twist angles (columns labelled $\alpha$), in which no interlayer interactions between the layers are considered. Note that no atomic relaxations occur in decoupled twisted bilauers. For a twist angle of $38.22\degree$, the phonon chiralities of the twisted bilayer are very similar to those of the monolayer and also to those of the decoupled system. As the twist angle approaches $60\degree$, two novel sets of chiral valley phonons emerge [indicated by the curly brackets in Fig.~\ref{fig2}(a)] in the twisted system which are absent in the monolayer and in the decoupled bilayer: one set with frequencies near 10~$\mathrm{cm^{-1}}$ and the other with frequencies near 160~$\mathrm{cm^{-1}}$. The frequencies of the first set soften as the twist angle approaches $60\degree$.

To understand the origin of these emergent chiral valley phonons, we plot the magnitude of the polarization vector in the moir\'e unit cell for 58$^\circ$, see Figs.~\ref{fig2}(b)-(e). Figs.~\ref{fig2}(b)-(c) show results for the two emergent valley phonons with the largest chiralities from the low-energy set (near $10~\mathrm{cm^{-1}}$). In particular, the mode with $6.2~\mathrm{cm^{-1}}$ has a chirality of $-0.56$ and is localized in the $\mathrm{B^{Se/Se}}$ stacking regions while the one with $9.6~\mathrm{cm^{-1}}$ has a chirality of 0.35 and is localized in the $\mathrm{B^{W/W}}$ stacking regions. Analyzing the contributions to the total chirality of these modes, we find that both layers and both W and Se atoms contribute. Figs.~\ref{fig2}(d)-(e) show the polarization vectors of the two valley phonons belonging to the second emergent set (with frequencies near $160~\mathrm{cm^{-1}}$) that have the largest chiralities. The mode with frequency $159.2~\mathrm{cm^{-1}}$ is localized in the $\mathrm{B^{Se/Se}}$ region while the one with frequency $159.4~\mathrm{cm^{-1}}$ is approximately localized around the $\mathrm{B^{W/W}}$ region. Interestingly, only the Se atoms contribute to the total chiralities of these modes.

The emergence of these novel chiral valley phonons can be explained by an inversion symmetry breaking on the moir\'e scale that occurs as the twist angle approaches $60\degree$. Near this twist angle, the moir\'e unit cell contains three high-symmetry stackings: 2H, $\mathrm{B^{W/W}}$, and
$\mathrm{B^{Se/Se}}$. Among these stackings 2H is most stable and for twist angles larger than $\sim~55\degree$ atomic relaxations result in a significant increase in the regions with 2H stacking as well as the formation of domain walls~\cite{Maityreconstruction2021} (see SI, Sec. III). It can be shown~\cite{Maityreconstruction2021} that this system can be mapped onto an equivalent system of two particles (with one particle representing $\mathrm{B^{W/W}}$ regions and the other $\mathrm{B^{Se/Se}}$ regions) that occupy different sublattices of a triangular lattice and are connected by springs. Since the energies of $\mathrm{B^{W/W}}$ and $\mathrm{B^{Se/Se}}$ regions are different, the two particles have different masses and therefore the effective system lacks inversion symmetry. This in turn results in the emergence of novel chiral phonon modes. In contrast, the twisted bilayer near $0\degree$ features three high-symmetry stackings: AA, $\mathrm{B^{Se/W}}$, and
$\mathrm{B^{W/Se}}$. Among these stackings $\mathrm{B^{Se/W}}$ and
$\mathrm{B^{W/Se}}$ are the most stable stackings. This system can be mapped onto a set of particles (representing $\mathrm{AA}$ regions) which occupy the sites of a triangular lattice and are connected by springs. This effective system exhibits inversion symmetry and therefore no emergent chiral phonon modes are observed.

Our analysis shows that the polarization vectors of chiral phonons near $0^\circ$ and $60^\circ$ are qualitatively
different from the those of the monolayer phonons. This suggests that commonly used approaches for calculating phonon properties in twisted bilayers that are based on a zone folding procedure of monolayer results are unreliable ~\cite{Linmoire2018, Parzefallmoire2021}.

\begin{figure}[!ht]
\centering
\includegraphics[width=0.5\textwidth]{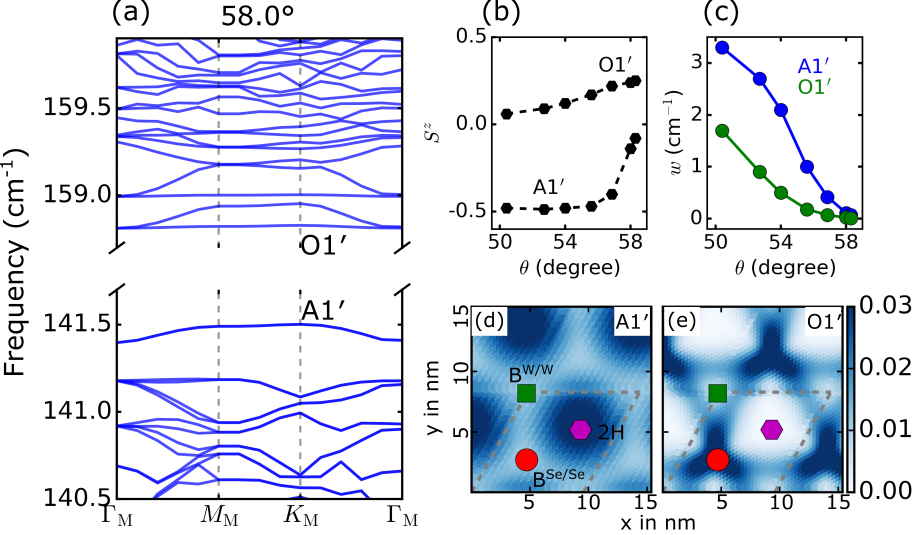}
\caption{(a) The phonon band structure of twisted bilayer $\mathrm{WSe_{2}}$ at $58^\circ$ in the vicinity of the energy gap between acoustic and optical modes. (b) Chirality $S^{z}$ (b) and band width (c) of flat acoustic ($\mathrm{A1^\prime}$) and optical ($\mathrm{O1^\prime}$) phonon bands as function of twist angle near $60^\circ$. (d) and (e): Polarization vectors of  $\mathrm{A1^\prime}$ and $\mathrm{O1^\prime}$ modes at a twist angle of
$58^\circ$. Different stackings are indicated by colored symbols and the moir\'{e} unit cell is outlined using dashed gray lines. In (d), the displacements of W atoms in the top layer is shown, while in (e) that of Se atoms in the top layer was used.}
\label{fig3}
\end{figure}

\textit{Phononic band flattening in twisted bilayer $\mathrm{WSe_{2}}$.} Figure~\ref{fig3}(a) demonstrates the emergence of flat phonon bands in the vicinity of the energy gap between acoustic and optical modes for a $58\degree$ twisted bilayer. This phononic band gap is inherited from the monolayer which has an energy gap of 17.9~$\mathrm{cm^{-1}}$ between the acoustic and optical phonon modes with the highest-energy acoustic mode being located at $K$ and the lowest-energy optical mode at the $\Gamma$ point of the BZ. The highest-energy acoustic mode at $K$ is non-degenerate and chiral with $S^{z}=-0.5$, while the lowest-energy optical modes at $\Gamma$ are degenerate and achiral. Fig.~\ref{fig3}(b) shows that the chirality associated with the highest-energy acoustic mode at $K_{\mathrm{M}}$ (denoted as $\mathrm{A1^\prime}$) changes dramatically and becomes achiral as the twist angle approaches $60^\circ$. This change in chirality is caused by the localization of the phonon mode in regions with inversion-symmetric 2H stacking (see discussion below). In contrast, the lowest-energy optical mode at $K_{\mathrm{M}}$ (denoted as $\mathrm{O1^\prime}$) becomes chiral as the twist angle approaches $60^\circ$. Fig.~\ref{fig3}(c) shows the bandwidths of the A1$^\prime$ and O1$^\prime$ modes and demonstrates that these bands become extremely flat as the twist angle approaches $60^\circ$ and the size of the moir\'e cell increases. Similar to the electronic bandwidths in twisted transition metal dichalcogenide bilayers, the flattening does not occur at a specific magic angle, unlike the case of twisted bilayer graphene~\cite{Bistritzermoire2011}. Interestingly, the widths of the flat phonon bands are significantly smaller than the width of the flat electron bands at the same twist angle. For example, at a twist angle of 56.5$^\circ$ of the twisted bilayer of TMDs, the width of the electronic valence band is 5 meV~\cite{Naikultraflat2018} and the width of the largest acoustic phonon band is 0.1 meV.

The flattening of the phonon bands is a consequence of both zone folding and the localization of phonons in real space (see SI, Sec. IV for details). Figs.~\ref{fig3}(d) and (e) show the polarization vectors (at $K_{\mathrm{M}}$) of the A1$^\prime$ and O1$^\prime$ modes for a twist angle of $58^\circ$. The A1$^\prime$ mode is localized in regions of 2H stacking, while O1$^\prime$ is localized in $\mathrm{B^{Se/Se}}$ regions as well as the domain walls. It is interesting to note that the localization of the highest acoustic phonons is strikingly similar to that of the highest electronic valence states in these systems~\cite{kunduflat2021}. Similar band flattening is also observed for twist angles near $0^\circ$ (see SI\cite{SI}, Sec. IV), where the highest-energy acoustic mode is localized in the $\mathrm{B^{W/Se}/B^{Se/W}}$ stacking region and the lowest-energy optical mode is localized in regions of $\mathrm{AA}$ stacking and domain walls. 

The distinct localization of the acoustic and optical
modes can be explained by analyzing the frequencies of these modes in the various untwisted bilayers with high-symmetry stacking arrangements. For example, the A1$^\prime$ mode originates from the $K$ point of the monolayer BZ. For the untwisted bilayers, the phonon frequencies at $K$ are 141.08~cm$^{-1}$ for $\mathrm{B^{Se/Se}}$ stacking, 141.1~cm$^{-1}$ for $\mathrm{B^{W/W}}$ stacking and 141.8~cm$^{-1}$ for $\mathrm{2H}$ stacking. Therefore, the highest-energy acoustic mode localizes on the $\mathrm{2H}$ regions as the moir\'e unit cell increases in size.

\textit{Chiral valley phonons in strain-engineered moir\'{e} materials and heterostructures.} Besides twisting, there are two approaches to create a moir\'e pattern in bilayer systems: (i) by applying strain to only one of two identical layers and (ii) by stacking two different 2D materials on top of each other. We
investigate the existence of chiral valley phonons in both cases. In contrast to the twisted bilayer at small twist angles, we find that the strain-engineered bilayer exhibits chiral phonons in the $K_{\mathrm{M}}$ valley (see Sec.~V of the SI for a summary of these modes and more details~\cite {SI}). In addition, emergent chiral phonon modes with frequencies of $\sim~10~\mathrm{cm^{-1}}$ and $\sim~156~\mathrm{cm^{-1}}$ are found. Finally, we also observe emergent chiral phonon modes in a $\mathrm{WSe_{2}/MoSe_{2}}$ heterostructure at a twist angle of $3.14^\circ$ (see SI~\cite{SI}, Sec. V).

\textit{Summary.} In this paper, we have demonstrated the presence of chiral phonons in twisted bilayer WSe$_2$ as well as other moir\'e materials and studied their properties. We have found that the phonon chirality depends on the twist angle with systems near $0^\circ$ exhibiting qualitatively different chiralities than systems near $60^\circ$. Very close to $60^\circ$, we observe emergent chiral modes as well as flat phonon bands. The predicted chiral properties of phonons in twisted bilayer materials can be measured with helicity-resolved Raman spectroscopy~\cite{Chenchiral2018,Tatsumiconservation2018}. While the resolution of such techniques is typically not high enough to access individual modes in systems near 0$^\circ$ and 60$^\circ$, they can potentially measure the k-point and helicity-resolved phonon density of states (see SI, Sec. VII). Future work should investigate the scattering of chiral phonons with other quasiparticles, such as electrons, excitons, and photons and the effects of such scattering processes on the electronic and optical properties of moir\'e materials.

\begin{acknowledgments}
\textit{Acknowledgments} This project has received funding from the European Union's Horizon 2020 research and innovation programme under the Marie Sk\l{}odowska-Curie grant agreement No 101028468. The authors acknowledge support from the Thomas Young Centre under grant TYC-101, and discussions with Rup Chowdhury, Nikita Tepliakov and Saurabh Srivastav.  
\end{acknowledgments}

\clearpage 
\newpage 

\large{\textbf{Supplementary Information (SI) : \\ Chiral valley phonons and flat phonon bands in moir\'{e} materials}}
\normalsize 
\section{I: Simulation details}

\subsection{Generation of structures}
\paragraph{Moir\'{e} pattern due to twist:}
The commensurate twist angles that give rise to the least-area moir\'{e} unit-cell for twisted bilayer of $\mathrm{WSe_{2}}$ are given by, 
\begin{equation}
\cos(\theta_{i}) = \frac{3i^{2} + 3i + 1/2}{3i^2 + 3i + 1}
\end{equation}
for an integer $i$ \cite{Lopesgraphene2007}. One of the set of
moir\'{e} lattice vectors are- ${\bf A_1}=i{\bf a_1} + (i+1){\bf
a_2}$ and ${\bf A_2}=-(i+1){\bf a_1} + (2i+1){\bf a_2}$ with ${\bf
a_1}=(1/2, \sqrt{3}/2)a_0$ and ${\bf a_2}=(-1/2, \sqrt{3}/2)a_0$,
where $a_0$ is the lattice constant of singe layer
$\mathrm{WSe_{2}}$. For twist angles ($\theta_i$) close to
$0^\circ$, we construct the moir\'{e} unit-cell starting from AA
stacking using TWISTER \cite{Naiktwister2021}. The high-symmetry
stackings present in the moir\'{e} patterns are- $\mathrm{AA}$
(with W and Se of the top layer are directly above W and Se of
bottom layer), $\mathrm{B^{W/Se}}$ (Bernal stacking with W of the
top layer directly above Se of bottom layer), and
$\mathrm{B^{Se/W}}$ (Bernal stacking with Se of the top layer
directly above W of bottom layer). These stackings are sometimes
referred to as AA, AB, and BA, as well. For twist angles
($60^\circ-\theta_i$) close to $60^\circ$, we construct the
moir\'{e} unit-cell starting from 2H stacking. The high-symmetry
stackings present in the moir\'{e} patterns are- $\mathrm{2H}$
(with W and Se of the top layer are directly above Se and W of
bottom layer), $\mathrm{B^{Se/Se}}$ (Bernal stacking with Se of the
top layer directly above Se of bottom layer), and
$\mathrm{B^{W/W}}$ (Bernal stacking with W of the top layer
directly above W of bottom layer). These stackings are sometimes
referred to as AA$^\prime$, A$^\prime$B, and AB$^\prime$,
respectively. 

\begin{table}[!ht]
\centering 
\begin{tabular}{|m{3cm}|m{3cm}|m{3.5cm}|}
\hline
Twist angles & Number of atoms & Moir\'{e} length (in \AA) \\
\hline
21.78$^\circ$ & 42 & 8.8 \\
\hline 
9.4$^\circ$ & 222 & 20.4 \\
\hline
7.3$^\circ$ & 366 & 26.1 \\
\hline
6$^\circ$ & 546 & 31.9 \\
\hline
4.41$^\circ$ & 1626 & 54.6 \\
\hline 
3.14$^\circ$ & 1986 & 60.4 \\
\hline 
2$^\circ$ & 4902 & 94.9 \\
\hline 
38.22$^\circ$ & 42 & 8.8 \\
\hline
50.6$^\circ$ & 222 & 20.4 \\
\hline
52.7$^\circ$ & 366 & 26.1 \\
\hline
54$^\circ$ & 546 & 31.9 \\
\hline
55.59$^\circ$ & 1626 & 54.6 \\
\hline  
56.86$^\circ$ & 1986 & 60.4 \\
\hline 
58$^\circ$ & 4902 & 94.9 \\
\hline 
58.31$^\circ$ & 6846 & 113.3 \\
\hline 
\end{tabular}
\caption{Moir\'{e} patterns of twisted bilayer $\mathrm{WSe_{2}}$ studied in this work. }
\end{table}

\paragraph{Moir\'{e} pattern due to strain}
We use the unit-cell lattice vectors of single layer of $\mathrm{WSe_{2}}$ as ${\bf a_{1}}=a_0(1, 0, 0)$, ${\bf a_{2}}=a_0(1/2, \sqrt{3}/2, 0)$, and ${\bf a_3}=a_0(0, 0, 30)$, where $a_0$ is the lattice constant. We can create the moir\'{e} patterns in two ways: by straining only one of the layer of $\mathrm{WSe_{2}}$, and by straining both the layers. We use the first approach in this work. We apply a tensile strain to the bottom layer. We discuss the details of this approach below. The periodicity of the strained bilayer is computed in the following manner:
\begin{equation}
\begin{bmatrix}
m & 0 \\
0 & m
\end{bmatrix}
\begin{pmatrix}
{\bf a_{1}} \\
{\bf a_{2}}
\end{pmatrix}
=
\begin{bmatrix}
n & 0 \\
0 & n
\end{bmatrix}
\begin{pmatrix}
{\bf a_{1}^{b}} \\
{\bf a_{2}^{b}}
\end{pmatrix}
\end{equation} 

In the case of 3.3 \% biaxial strain, we use $m=30$, and $n=29$ starting from AA stacking configuration. The moir\'{e} lattice constant of the strained bilayer is 98.9 \AA\ and contains 5223 atoms.

\paragraph{Moir\'{e} heterostructure}
We replace the top layer of bilayer of $\mathrm{WSe_{2}}$ by $\mathrm{MoSe_{2}}$ and construct moir\'{e} lattice by rotating the layer. Since the experimental lattice constants of these two materials are similar, we simply replace one of the $\mathrm{WSe_{2}}$ layer by $\mathrm{MoSe_{2}}$ in the twisted bilayer $\mathrm{WSe_{2}}$ and create the moir\'{e} heterostructure.  

\subsection{Structural relaxations}
The moir\'{e} patterns are relaxed using the LAMMPS package with the Stillinger-Weber \cite{Arashvalidation2018}, and Kolmogorov-Crespi \cite{Naikkolmogorov2019} potentials to capture the intralayer and interlayer interactions of the twisted bilayer of $\mathrm{WSe_{2}}$, respectively. The Kolmogorov-Crespi parameters used in this work can correctly reproduce the interlayer binding energy landscape, obtained using density functional theory. The atomic relaxations produced using these parameters are in excellent agreement with relaxations performed using density functional theory. The intralayer Stillinger-Weber potential quite accurately captures the full phonon dispersion, in particular the acoustic modes. The accurate potentials used in our simulations are expected to capture the phonon renormalization of the moir\'{e} materials. At first, we relax the simulation box to ensure the in-plane pressure is as small as possible. Next, we relax the atoms within a fixed simulation box with the force tolerance of $10^{-5}$ eV/\AA\ for any atom along any direction.

\subsection{Force constants calculations}

For monolayer $\mathrm{WSe_{2}}$, we use a $6\times6\times1$ supercell to compute the force constants. For large twist angles of twisted bilayer $\mathrm{WSe_{2}}$, we use a $2\times2\times1$ supercell to compute the force constants. The moir\'{e} patterns contain thousands of atoms for small twist angles. Therefore, we use $1\times1\times1$ supercell to compute the force-constants. 

\clearpage
\newpage

\section{II : Chiral phonons in 2D materials}
\subsection{Absence of chiral valley-phonons in graphene}
\begin{figure}[!ht]
\centering
\includegraphics[width=0.6\textwidth]{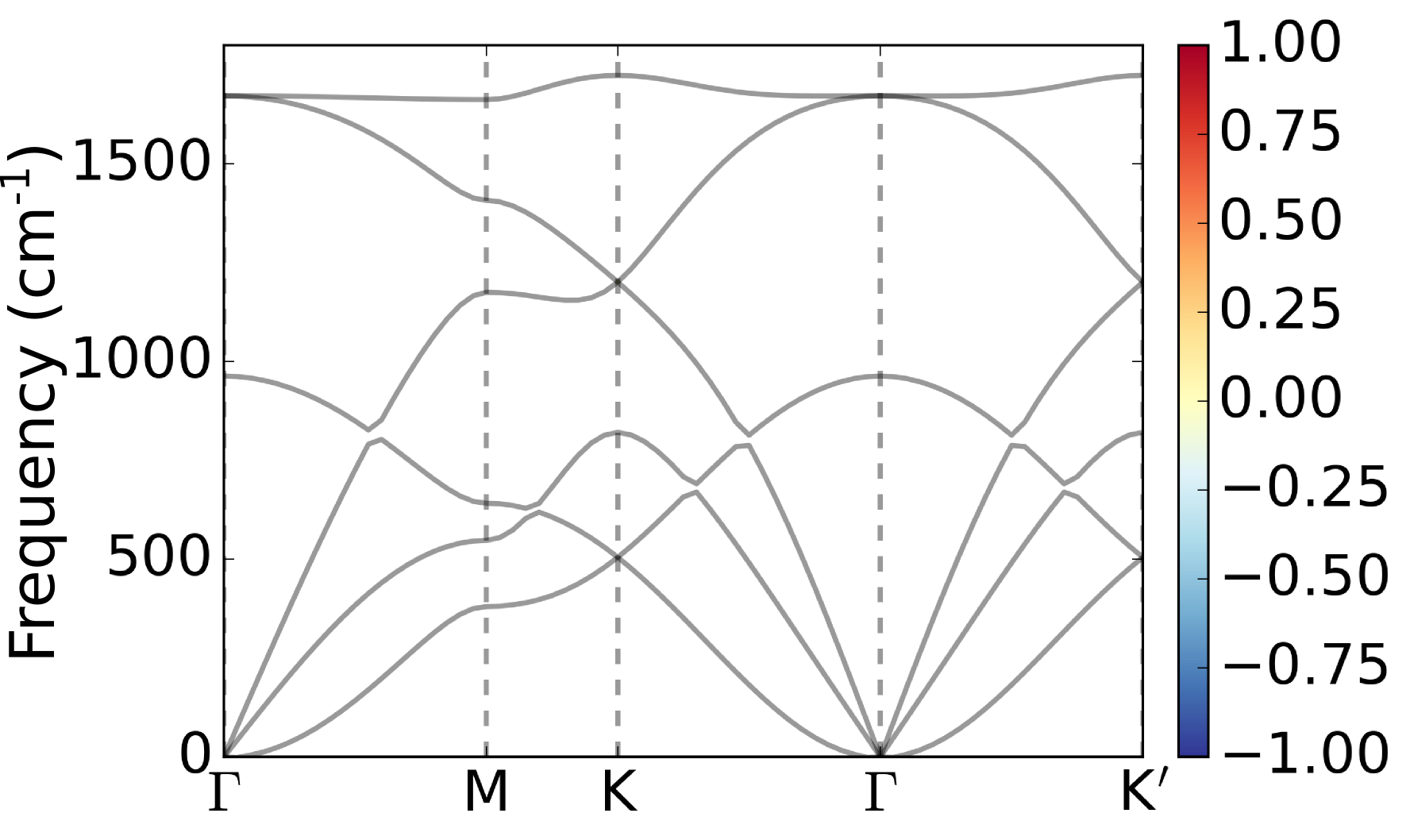}
\caption{Chirality of phonons in single layer of graphene. The solid lines represent the phonon frequencies, and hexagons (absent in the figure as $S^{z}=0$) represent the computed $S_{z}$ values after summing over all the basis atoms. There are no non-degenerate chiral modes in single layer of graphene due to inversion symmetry.}  
\end{figure}

\begin{table}[!ht]
\centering 
\begin{tabular}{|m{2cm}|m{2cm}|m{2cm}|m{2cm}|}
\hline
$\omega$ & $S^{z}$ & $S^{z}_{C1}$ & $S^{z}_{C{2}}$ \\
(in $\mathrm{cm^{-1}}$) &  &  &  \\
\hline
505.2 & 0.0 & 0.0 & 0.0 \\
\hline 
505.2 & 0.0 & 0.0 & 0.0 \\
\hline 
821.1 & 0.0 & 0.5 & -0.5 \\
\hline 
1201.5 & 0.0 & -0.5  & 0.5 \\ 
\hline 
1201.5 & 0.0 & -0.5  & 0.5 \\
\hline 
1724.5 & 0.0 & 0.5 & -0.5 \\
\hline 
\end{tabular}
\caption{Phonon modes, total chirality, and chirality of the each types of atoms in the case of single-layer of graphene. Within the unit-cell, two atoms can have opposite chirality, but the total chirality is 0.}
\end{table}

\clearpage 
\newpage 
\begin{tikzpicture}
\def\x1{0}; \def\y1{0};\def\z{3}
\def\rx{1.5}; \def\ry{0.4};

\draw [very thick] (\x1-2, \y1-1.5-\ry)--(\x1,\y1-\ry)--(\x1+2.5, \y1-\ry)--(\x1+4.5, \y1-\ry-1.5)--(\x1+2.5, -\y1-\ry-\z)--(\x1,\y1-\ry-\z)--(\x1-2, \y1-1.5-\ry);

\draw[blue,fill=blue] (\x1-2, \y1-1.5-\ry) circle (\ry*0.5 cm);
\draw[blue,fill=blue] (\x1,\y1-\ry) circle (\ry*0.5 cm);
\draw[blue,fill=blue] (\x1+2.5, \y1-\ry) circle (\ry*0.5 cm);
\draw[blue,fill=blue] (\x1+4.5, \y1-\ry-1.5) circle (\ry*0.5 cm);
\draw[blue,fill=blue] (\x1+2.5, -\y1-\ry-\z) circle (\ry*0.5 cm);
\draw[blue,fill=blue] (\x1,\y1-\ry-\z) circle (\ry*0.5 cm);

\draw [-latex, ultra thick, rotate=0] (\x1-2, \y1-1.5-\ry)--(\x1-2,\y1-\ry-1.5+1);
\draw [-latex, ultra thick, rotate=0] (\x1,\y1-\ry)--(\x1,\y1-\ry+0.5);
\node at (\x1 + 1.25,-\z/2) {$\omega=505.2$ cm$^{-1}$ };
-------------------------------

\def\x1{9}; \def\y1{0};\def\z{3}
\def\rx{1.5}; \def\ry{0.4};
\draw [very thick] (\x1-2, \y1-1.5-\ry)--(\x1,\y1-\ry)--(\x1+2.5, \y1-\ry)--(\x1+4.5, \y1-\ry-1.5)--(\x1+2.5, -\y1-\ry-\z)--(\x1,\y1-\ry-\z)--(\x1-2, \y1-1.5-\ry);

\draw[blue,fill=blue] (\x1-2, \y1-1.5-\ry) circle (\ry*0.5 cm);
\draw[blue,fill=blue] (\x1,\y1-\ry) circle (\ry*0.5 cm);
\draw[blue,fill=blue] (\x1+2.5, \y1-\ry) circle (\ry*0.5 cm);
\draw[blue,fill=blue] (\x1+4.5, \y1-\ry-1.5) circle (\ry*0.5 cm);
\draw[blue,fill=blue] (\x1+2.5, \y1-\ry-\z) circle (\ry*0.5 cm);
\draw[blue,fill=blue] (\x1,\y1-\ry-\z) circle (\ry*0.5 cm);

\draw [-latex, ultra thick, rotate=0] (\x1-2, \y1-1.5-\ry)--(\x1-2,\y1-\ry-1.5+0.5);
\draw [-latex, ultra thick, rotate=0] (\x1,\y1-\ry)--(\x1,\y1-\ry+1);
\node at (\x1 + 1.25,-\z/2) {$\omega=505.2$ cm$^{-1}$ };

\def\x1{0}; \def\y1{-6};\def\z{3}
\def\rx{1.5}; \def\ry{0.4};

\draw [very thick] (\x1-2, \y1-1.5-\ry)--(\x1,\y1-\ry)--(\x1+2.5, \y1-\ry)--(\x1+4.5, \y1-\ry-1.5)--(\x1+2.5, \y1-\ry-\z)--(\x1,\y1-\ry-\z)--(\x1-2, \y1-1.5-\ry);

\draw[blue,fill=blue] (\x1-2, \y1-1.5-\ry) circle (\ry*0.5 cm);
\draw[blue,fill=blue] (\x1,\y1-\ry) circle (\ry*0.5 cm);
\draw[blue,fill=blue] (\x1+2.5, \y1-\ry) circle (\ry*0.5 cm);
\draw[blue,fill=blue] (\x1+4.5, \y1-\ry-1.5) circle (\ry*0.5 cm);
\draw[blue,fill=blue] (\x1+2.5, \y1-\ry-\z) circle (\ry*0.5 cm);
\draw[blue,fill=blue] (\x1,\y1-\ry-\z) circle (\ry*0.5 cm);

\draw [-latex, ultra thick, rotate=0] (\x1-2, \y1-1.5) arc [start angle=90, end angle=430, x radius=\rx cm, y radius=\ry cm];
\draw [latex-, ultra thick, rotate=0] (\x1,\y1) arc [start angle=90, end angle=430, x radius=\rx cm, y radius=\ry cm];
\node at (\x1 + 1.25, \y1 -\z/2) {$\omega=821.1$ cm$^{-1}$ };

\def\x1{9}; \def\y1{-6};\def\z{3}
\def\rx{1.5}; \def\ry{0.4};

\draw [very thick] (\x1-2, \y1-1.5-\ry)--(\x1,\y1-\ry)--(\x1+2.5, \y1-\ry)--(\x1+4.5, \y1-\ry-1.5)--(\x1+2.5, \y1-\ry-\z)--(\x1,\y1-\ry-\z)--(\x1-2, \y1-1.5-\ry);

\draw[blue,fill=blue] (\x1-2, \y1-1.5-\ry) circle (\ry*0.5 cm);
\draw[blue,fill=blue] (\x1,\y1-\ry) circle (\ry*0.5 cm);
\draw[blue,fill=blue] (\x1+2.5, \y1-\ry) circle (\ry*0.5 cm);
\draw[blue,fill=blue] (\x1+4.5, \y1-\ry-1.5) circle (\ry*0.5 cm);
\draw[blue,fill=blue] (\x1+2.5, \y1-\ry-\z) circle (\ry*0.5 cm);
\draw[blue,fill=blue] (\x1,\y1-\ry-\z) circle (\ry*0.5 cm);

\draw [latex-, ultra thick, rotate=0] (\x1-2, \y1-1.5) arc [start angle=90, end angle=430, x radius=\rx cm, y radius=\ry cm];
\draw [-latex, ultra thick, rotate=0] (\x1,\y1) arc [start angle=90, end angle=430, x radius=\rx cm, y radius=\ry cm];
\node at (\x1 + 1.25, \y1 -\z/2) {$\omega=1201.5$ cm$^{-1}$ };

\def\x1{0}; \def\y1{-12};\def\z{3}
\def\rx{1.5}; \def\ry{0.4};

\draw [very thick] (\x1-2, \y1-1.5-\ry)--(\x1,\y1-\ry)--(\x1+2.5, \y1-\ry)--(\x1+4.5, \y1-\ry-1.5)--(\x1+2.5, \y1-\ry-\z)--(\x1,\y1-\ry-\z)--(\x1-2, \y1-1.5-\ry);

\draw[blue,fill=blue] (\x1-2, \y1-1.5-\ry) circle (\ry*0.5 cm);
\draw[blue,fill=blue] (\x1,\y1-\ry) circle (\ry*0.5 cm);
\draw[blue,fill=blue] (\x1+2.5, \y1-\ry) circle (\ry*0.5 cm);
\draw[blue,fill=blue] (\x1+4.5, \y1-\ry-1.5) circle (\ry*0.5 cm);
\draw[blue,fill=blue] (\x1+2.5, \y1-\ry-\z) circle (\ry*0.5 cm);
\draw[blue,fill=blue] (\x1,\y1-\ry-\z) circle (\ry*0.5 cm);

\draw [latex-, ultra thick, rotate=0] (\x1-2, \y1-1.5) arc [start angle=90, end angle=430, x radius=\rx cm, y radius=\ry cm];
\draw [-latex, ultra thick, rotate=0] (\x1,\y1) arc [start angle=90, end angle=430, x radius=\rx cm, y radius=\ry cm];
\node at (\x1 + 1.25, \y1 -\z/2) {$\omega=1201.5$ cm$^{-1}$ };

\def\x1{9}; \def\y1{-12};\def\z{3}
\def\rx{1.5}; \def\ry{0.4};

\draw [very thick] (\x1-2, \y1-1.5-\ry)--(\x1,\y1-\ry)--(\x1+2.5, \y1-\ry)--(\x1+4.5, \y1-\ry-1.5)--(\x1+2.5, \y1-\ry-\z)--(\x1,\y1-\ry-\z)--(\x1-2, \y1-1.5-\ry);

\draw[blue,fill=blue] (\x1-2, \y1-1.5-\ry) circle (\ry*0.5 cm);
\draw[blue,fill=blue] (\x1,\y1-\ry) circle (\ry*0.5 cm);
\draw[blue,fill=blue] (\x1+2.5, \y1-\ry) circle (\ry*0.5 cm);
\draw[blue,fill=blue] (\x1+4.5, \y1-\ry-1.5) circle (\ry*0.5 cm);
\draw[blue,fill=blue] (\x1+2.5, \y1-\ry-\z) circle (\ry*0.5 cm);
\draw[blue,fill=blue] (\x1,\y1-\ry-\z) circle (\ry*0.5 cm);

\draw [-latex, ultra thick, rotate=0] (\x1-2, \y1-1.5) arc [start angle=90, end angle=430, x radius=\rx cm, y radius=\ry cm];
\draw [latex-, ultra thick, rotate=0] (\x1, \y1) arc [start angle=90, end angle=430, x radius=\rx cm, y radius=\ry cm];
\node at (\x1 + 1.25, \y1 -\z/2) {$\omega=1724.5$ cm$^{-1}$ };

\node at (5, \y1 -2*\z) {Eigenvectors associated with phonon modes in graphene};
\node at (5, \y1 -2*\z-0.5) {The Blue circles denote the Carbon atoms};
\end{tikzpicture}

\clearpage 
\newpage 

\subsection{Presence of chiral non-degenerate valley phonons in monolayer $\mathrm{WSe_{2}}$}

\begin{figure}[!ht]
\centering
\includegraphics[width=0.6\textwidth]{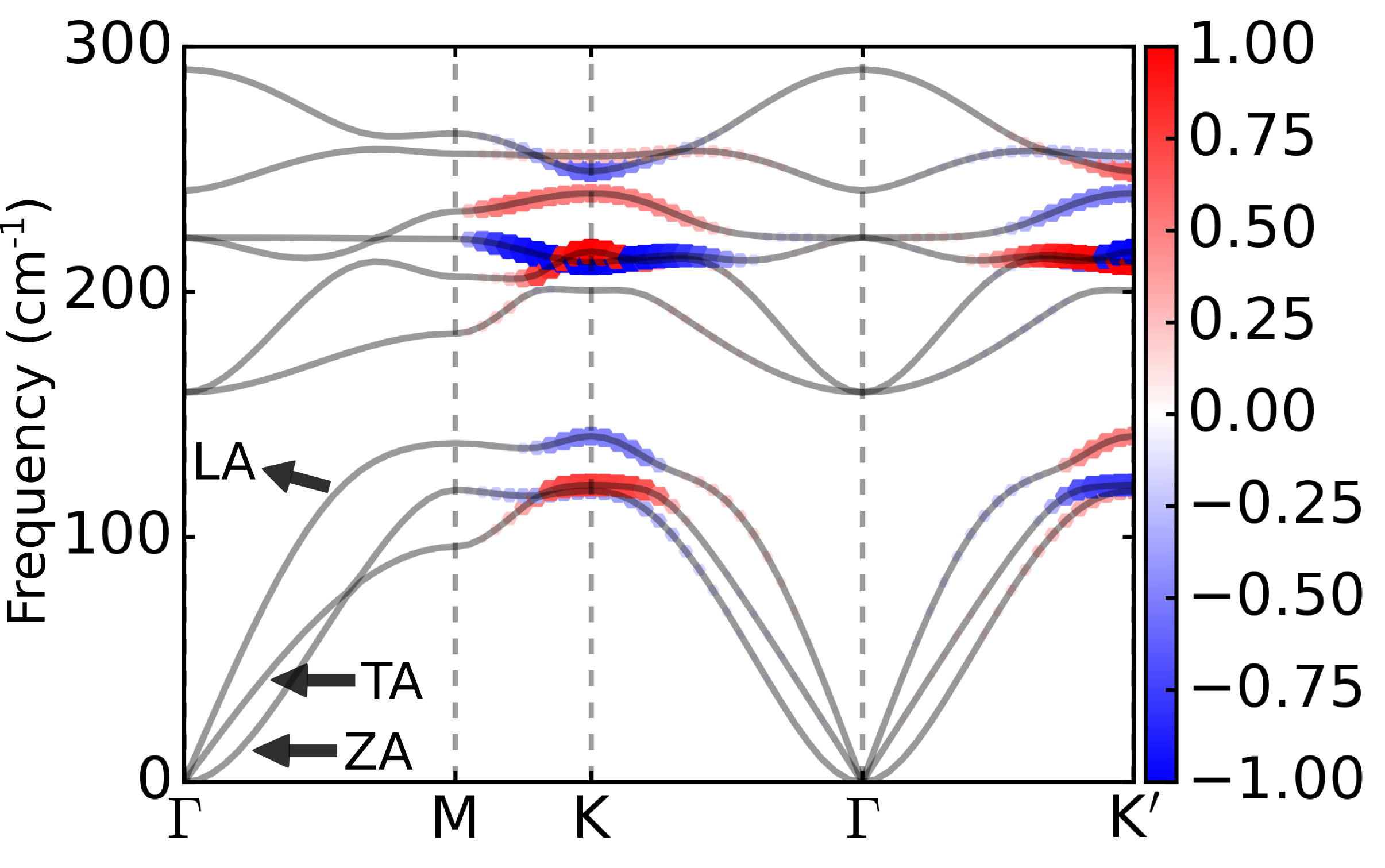}
\caption{Chirality of phonons in monolayer of $\mathrm{WSe_{2}}$. The solid lines represent the phonon frequencies, and hexagons represent the computed $S_{z}$ values after summing over all the basis atoms. There are multiple non-degenerate chiral modes in monolayer of $\mathrm{WSe_{2}}$ due to inversion symmetry breaking.}  
\end{figure}

\begin{table}[!ht]
\centering 
\begin{tabular}{|m{2cm}|m{2cm}|m{2cm}|m{2cm}|m{2cm}|}
\hline
$\omega$ & $S^{z}$ & $S^{z}_{W1}$ & $S^{z}_{Se{2}}$ & $S^{z}_{Se{3}}$  \\
(in $\mathrm{cm^{-1}}$) &  &  &  &  \\
\hline
118.9 & -0.38 & 0.0  & -0.19 & -0.19 \\
\hline
121.2 &  0.76 & 0.76 & 0.0   &  0.0 \\
\hline   
141.4 & -0.49 & -0.75 & 0.13 & 0.13 \\
\hline 
200.66 & 0.0  & 0.0  & 0.0 & 0.0 \\
\hline 
212.06 & -1.0 & 0.0 & -0.5 & -0.5 \\
\hline 
216.52 & 1.0 & 0.0 & 0.5 & 0.5 \\
\hline 
240.28 & 0.49 & -0.25 & 0.37 & 0.37 \\   
\hline 
249.2 & -0.62 & 0.0 & -0.31 & -0.31 \\
\hline 
255.38 & 0.24 & 0.24 & 0.0 & 0.0 \\
\hline  
\end{tabular}
\caption{Chirality associated with all the valley phonon modes (at $K$) in single layer of $\mathrm{WSe_{2}}$.  }
\label{structures}
\end{table}

\clearpage 
\newpage 

\begin{tikzpicture}
\def\x1{0}; \def\y1{0};\def\z{3}
\def\rx{1.5}; \def\ry{0.4};

\draw [latex-, ultra thick, rotate=0] (\x1 ,\y1) arc [start angle=90, end angle=430, x radius=\rx cm, y radius=\ry cm];
\draw[blue,fill=blue] (0,-\ry) circle (\ry*0.5 cm);

\draw [latex-, ultra thick, rotate=0] (\x1, {\y1-\z}) arc [start angle=90, end angle=430, x radius=\rx cm, y radius=\ry cm];
\draw[blue,fill=blue] (0, {-\ry - \z}) circle (\ry*0.5 cm);

\draw [-latex, ultra thick, rotate=0] (\x1-2, \y1-1.9)--(\x1-2,\y1-0.9);
\draw[red,fill=red] (\x1-2,\y1-1.9) circle (\ry*0.5 cm);

\draw [thick] (-2, -1.9)--(0,-0.4);
\draw [thick] (-2, -1.9)--(0, -3.4);

\node at (\x1 , \y1 -\z-1.2) {$\omega=118.9$ cm$^{-1}$ };

\def\x1{6}; \def\y1{0};\def\z{3}
\def\rx{1.5}; \def\ry{0.4};
\draw [-latex, ultra thick, rotate=0] (\x1, -\ry)--(\x1,{-\ry+1});
\draw[blue,fill=blue] (\x1,-\ry) circle (\ry*0.5 cm);

\draw [-latex, ultra thick, rotate=0] (\x1, -\ry-\z)--(\x1,{-\ry-\z+1});
\draw[blue,fill=blue] (\x1, {-\ry - \z}) circle (\ry*0.5 cm);

\draw [-latex, ultra thick, rotate=0] (\x1-2, {\y1-1.5}) arc [start angle=90, end angle=430, x radius=\rx cm, y radius=\ry cm];
\draw[red,fill=red] (\x1-2,\y1-1.5-\ry) circle (\ry*0.5 cm);

\draw [thick] (\x1-2, \y1-1.5-\ry)--(\x1,\y1-\ry);
\draw [thick] (\x1-2, \y1-1.5-\ry)--(\x1, \y1-\z-\ry);

\node at (\x1 , \y1 -\z-1.2) {$\omega=121.2$ cm$^{-1}$ };

\def\x1{12}; \def\y1{0};\def\z{3}
\def\rx{1.5}; \def\ry{0.4};
\draw [-latex, ultra thick, rotate=0] (\x1, {\y1-\z}) arc [start angle=90, end angle=430, x radius=\rx cm, y radius=\ry cm];
\draw[blue,fill=blue] (\x1,-\ry) circle (\ry*0.5 cm);

\draw [-latex, ultra thick, rotate=0] (\x1, {\y1}) arc [start angle=90, end angle=430, x radius=\rx cm, y radius=\ry cm];
\draw[blue,fill=blue] (\x1, {-\ry - \z}) circle (\ry*0.5 cm);

\draw [latex-, ultra thick, rotate=0] (\x1-2, {\y1-1.5}) arc [start angle=90, end angle=430, x radius=\rx cm, y radius=\ry cm];
\draw[red,fill=red] (\x1-2,\y1-1.5-\ry) circle (\ry*0.5 cm);

\draw [thick] (\x1-2, \y1-1.5-\ry)--(\x1,\y1-\ry);
\draw [thick] (\x1-2, \y1-1.5-\ry)--(\x1, \y1-\z-\ry);

\node at (\x1 , \y1 -\z-1.2) {$\omega=141.4$ cm$^{-1}$ };

\def\x1{0}; \def\y1{-6};\def\z{3}
\def\rx{1.5}; \def\ry{0.4};
\draw [-latex, ultra thick, rotate=0] (\x1, \y1-\ry)--(\x1,{\y1-\ry+1});
\draw[blue,fill=blue] (\x1,\y1-\ry) circle (\ry*0.5 cm);

\draw [-latex, ultra thick, rotate=0] (\x1, \y1-\ry-\z)--(\x1,{\y1-\ry+1-\z});
\draw[blue,fill=blue] (\x1, {\y1 -\ry - \z}) circle (\ry*0.5 cm);

\draw[red,fill=red] (\x1-2,\y1-1.5-\ry) circle (\ry*0.5 cm);

\draw [thick] (\x1-2, \y1-1.5-\ry)--(\x1,\y1-\ry);
\draw [thick] (\x1-2, \y1-1.5-\ry)--(\x1, \y1-\z-\ry);

\node at (\x1 , \y1 -\z-1.2) {$\omega=200.6$ cm$^{-1}$ };

\def\x1{6}; \def\y1{-6};\def\z{3}
\def\rx{1.5}; \def\ry{0.4};
\draw [latex-, ultra thick, rotate=0] (\x1, {\y1-\z}) arc [start angle=90, end angle=430, x radius=\rx cm, y radius=\ry cm];
\draw[blue,fill=blue] (\x1,\y1-\ry) circle (\ry*0.5 cm);

\draw [latex-, ultra thick, rotate=0] (\x1, {\y1}) arc [start angle=90, end angle=430, x radius=\rx cm, y radius=\ry cm];
\draw[blue,fill=blue] (\x1, {\y1-\ry - \z}) circle (\ry*0.5 cm);

\draw[red,fill=red] (\x1-2,\y1-1.5-\ry) circle (\ry*0.5 cm);

\draw [thick] (\x1-2, \y1-1.5-\ry)--(\x1,\y1-\ry);
\draw [thick] (\x1-2, \y1-1.5-\ry)--(\x1, \y1-\z-\ry);

\node at (\x1 , \y1 -\z-1.2) {$\omega=212$ cm$^{-1}$ };

\def\x1{12}; \def\y1{-6};\def\z{3}
\def\rx{1.5}; \def\ry{0.4};
\draw [-latex, ultra thick, rotate=0] (\x1, {\y1-\z}) arc [start angle=90, end angle=430, x radius=\rx cm, y radius=\ry cm];
\draw[blue,fill=blue] (\x1,\y1-\ry) circle (\ry*0.5 cm);

\draw [-latex, ultra thick, rotate=0] (\x1, {\y1}) arc [start angle=90, end angle=430, x radius=\rx cm, y radius=\ry cm];
\draw[blue,fill=blue] (\x1, {\y1-\ry - \z}) circle (\ry*0.5 cm);

\draw[red,fill=red] (\x1-2,\y1-1.5-\ry) circle (\ry*0.5 cm);

\draw [thick] (\x1-2, \y1-1.5-\ry)--(\x1,\y1-\ry);
\draw [thick] (\x1-2, \y1-1.5-\ry)--(\x1, \y1-\z-\ry);

\node at (\x1 , \y1 -\z-1.2) {$\omega=216.5$ cm$^{-1}$ };

\def\x1{0}; \def\y1{-12};\def\z{3}
\def\rx{1.5}; \def\ry{0.4};
\draw [-latex, ultra thick, rotate=0] (\x1, {\y1-\z}) arc [start angle=90, end angle=430, x radius=\rx cm, y radius=\ry cm];
\draw[blue,fill=blue] (\x1,\y1-\ry) circle (\ry*0.5 cm);

\draw [-latex, ultra thick, rotate=0] (\x1, {\y1}) arc [start angle=90, end angle=430, x radius=\rx cm, y radius=\ry cm];
\draw[blue,fill=blue] (\x1, {\y1-\ry - \z}) circle (\ry*0.5 cm);

\draw [latex-, ultra thick, rotate=0] (\x1-2, {\y1-1.5}) arc [start angle=90, end angle=430, x radius=\rx cm, y radius=\ry cm];
\draw[red,fill=red] (\x1-2,\y1-1.5-\ry) circle (\ry*0.5 cm);

\draw [thick] (\x1-2, \y1-1.5-\ry)--(\x1,\y1-\ry);
\draw [thick] (\x1-2, \y1-1.5-\ry)--(\x1, \y1-\z-\ry);

\node at (\x1 , \y1 -\z-1.2) {$\omega=240.3$ cm$^{-1}$ };

\def\x1{6}; \def\y1{-12};\def\z{3}
\def\rx{1.5}; \def\ry{0.4};
\draw [latex-, ultra thick, rotate=0] (\x1 ,\y1) arc [start angle=90, end angle=430, x radius=\rx cm, y radius=\ry cm];
\draw[blue,fill=blue] (\x1,\y1-\ry) circle (\ry*0.5 cm);

\draw [latex-, ultra thick, rotate=0] (\x1, {\y1-\z}) arc [start angle=90, end angle=430, x radius=\rx cm, y radius=\ry cm];
\draw[blue,fill=blue] (\x1, {\y1-\ry - \z}) circle (\ry*0.5 cm);

\draw [-latex, ultra thick, rotate=0] (\x1-2, \y1-1.9)--(\x1-2,\y1-0.9);
\draw[red,fill=red] (\x1-2,\y1-1.9) circle (\ry*0.5 cm);

\draw [thick] (\x1-2, \y1-1.5-\ry)--(\x1,\y1-\ry);
\draw [thick] (\x1-2, \y1-1.5-\ry)--(\x1, \y1-\z-\ry);
\node at (\x1 , \y1 -\z-1.2) {$\omega=249.2$ cm$^{-1}$ };

\def\x1{12}; \def\y1{-12};\def\z{3}
\def\rx{1.5}; \def\ry{0.4};
\draw [-latex, ultra thick, rotate=0] (\x1, \y1-\ry)--(\x1,{\y1-\ry+1});
\draw[blue,fill=blue] (\x1,\y1-\ry) circle (\ry*0.5 cm);

\draw [-latex, ultra thick, rotate=0] (\x1, \y1-\ry-\z)--(\x1,{\y1-\ry+1-\z});
\draw[blue,fill=blue] (\x1, {\y1-\ry - \z}) circle (\ry*0.5 cm);

\draw [-latex, ultra thick, rotate=0] (\x1-2, {\y1-1.5}) arc [start angle=90, end angle=430, x radius=\rx cm, y radius=\ry cm];
\draw[red,fill=red] (\x1-2,\y1-1.9) circle (\ry*0.5 cm);

\draw [thick] (\x1-2, \y1-1.5-\ry)--(\x1,\y1-\ry);
\draw [thick] (\x1-2, \y1-1.5-\ry)--(\x1, \y1-\z-\ry);

\node at (\x1 , \y1 -\z-1.2) {$\omega=255.4$ cm$^{-1}$ };

\node at (5, \y1 -2*\z) {Eigenvectors associated with phonon modes in single layer $\mathrm{WSe_{2}}$};
\node at (5, \y1 -2*\z-0.5) {The Blue and red circles denote the Se and W atoms, respectively.};

\end{tikzpicture}
\clearpage 
\newpage 

\section{III : Interlayer separation for twist angles close to $60^\circ$ and $0^\circ$}
\begin{figure}[!ht]
\centering
\includegraphics[width=0.5\textwidth]{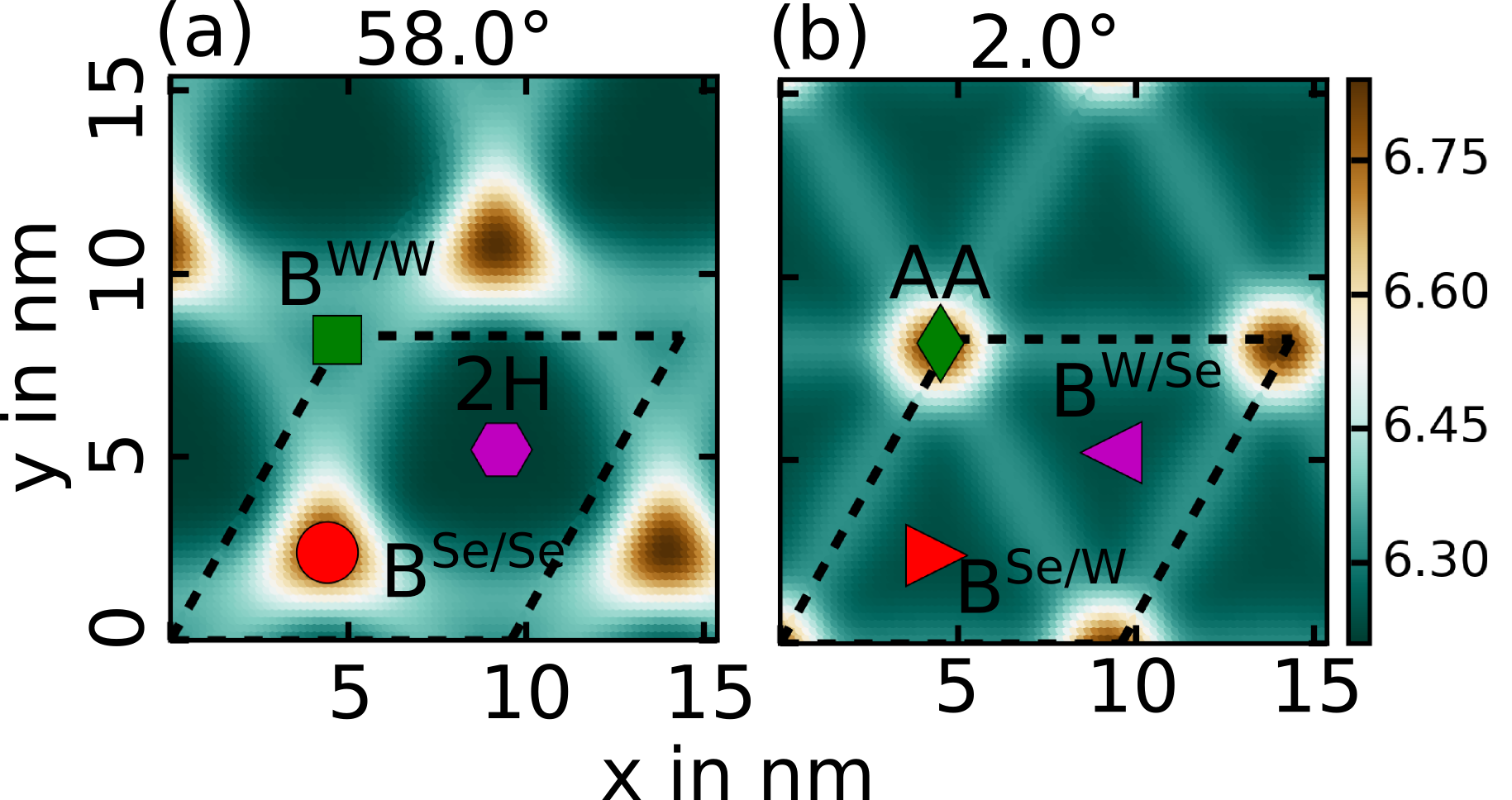}
\caption{(a),(b): Significant structural relaxations and inversion symmetry breaking at the moir\'{e} scale illustrated using interlayer separation for twist angles of $58^\circ$ and $2^\circ$, respectively. The associated color bar denotes the magnitude of interlayer separation and is in \AA. The moir\'{e} unit cell and stackings are marked.}  
\end{figure}

\clearpage 
\newpage 

\section{IV : Flat phonon bands} 
\begin{figure}[!ht]
\centering
\includegraphics[width=0.7\textwidth]{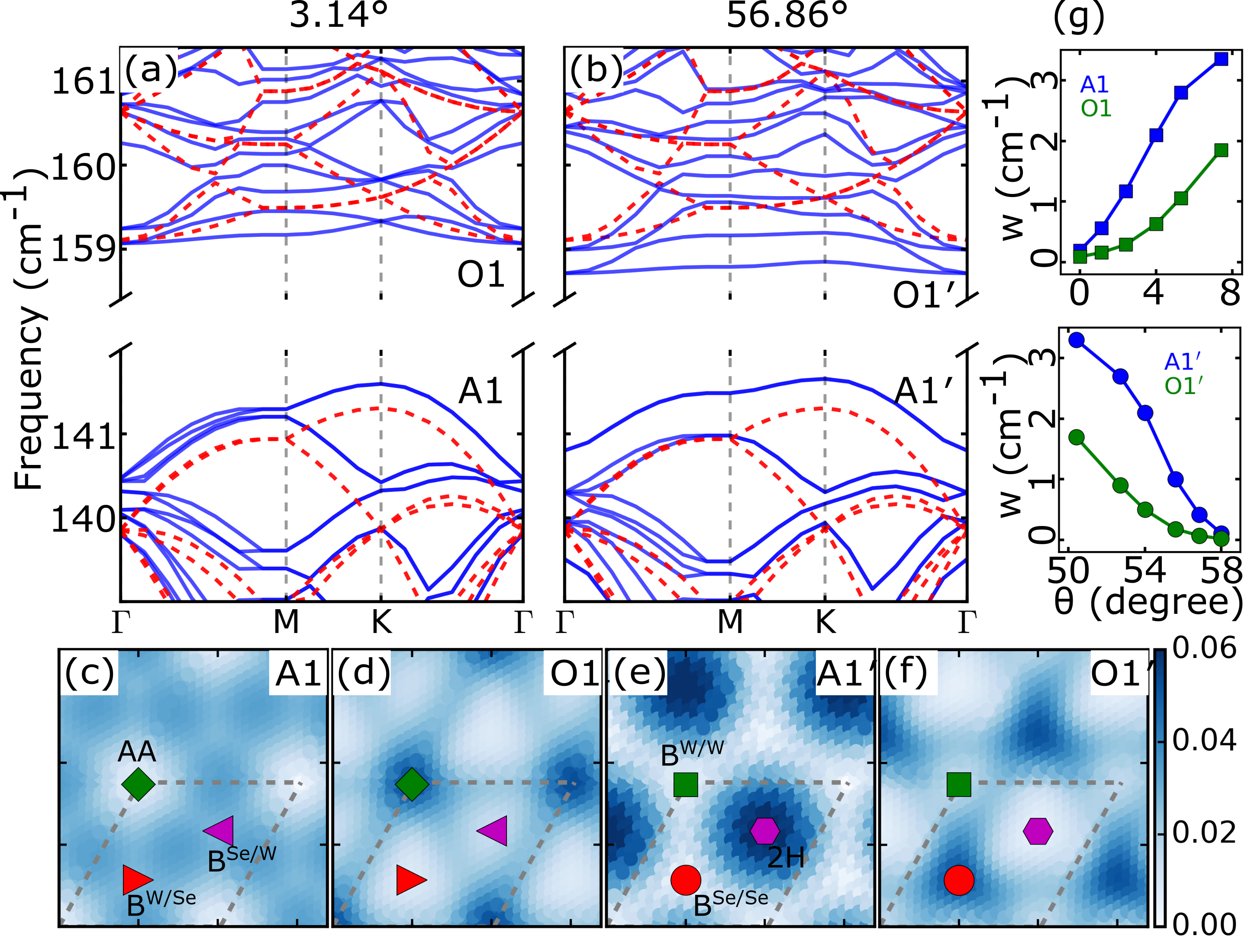}
\caption{(a), (b) Evidence of band-flattening for a twist angle of 3.14$^\circ$ and $56.86^\circ$ twisted bilayer $\mathrm{WSe_{2}}$ near the acoustic-optic phonon band gap. The blue lines denote the calculations with twisted bilayers and the red dashed lines denote calculations with decoupled twisted bilayers. Importantly, the band flattens and degeneracies in the band-structure of the decoupled systems are significantly reduced. (c),(d) ((e),(f)) Absolute value of the polarization vectors at $\Gamma_{\mathrm{M}}$ of  $\mathrm{A1}$ and $\mathrm{O1}$ modes at a twist angle of $3.14^\circ$ ($56.86^\circ$). Different stackings are indicated by colored symbols and the moir\'{e} unit cell is outlined using dashed gray lines. In (c),(e) the displacements of W atoms in the top layer is shown, while in (d),(f) that of Se atoms in the top layer was used. (g) The band-flattening is indicated by the reduction of band-width with twist angles. }  
\end{figure}

\begin{table}[h]
\centering
\begin{tabular}{  c c c c c }
\hline
 & \multicolumn{4}{c}{Bandwidth in cm$^{-1}$} \\
 & \multicolumn{2}{c}{Twisted bilayer} & \multicolumn{2}{c}{Decoupled twisted bilayer} \\
 \hline\hline 
Twist angle & A1 & O1 & A1 & O1 \\
\hline 
55.59$^\circ$ & 1.029 & 0.185 & 1.321 & 0.327 \\
56.86$^\circ$ & 0.427 & 0.069 & 0.672 & 0.110 \\

\hline
\end{tabular}
\caption{Comparison of bandwidths in the twisted bilayers that include all the relaxation effects and in the decoupled twisted bilayers due to zone-folding for $55.59^\circ$ and $56.86^\circ$ twisted bilayers. Note that no atomic relaxations occur in decoupled twisted bilayers.}
\label{bandwidth}
\end{table}

\clearpage
\newpage 

\section{V : Chiral valley phonons in strain engineered moir\'{e} and $\mathrm{MoSe_{2}/WSe_{2}}$ heterostructure} 
\begin{figure}[!ht]
\centering
\includegraphics[width=0.7\textwidth]{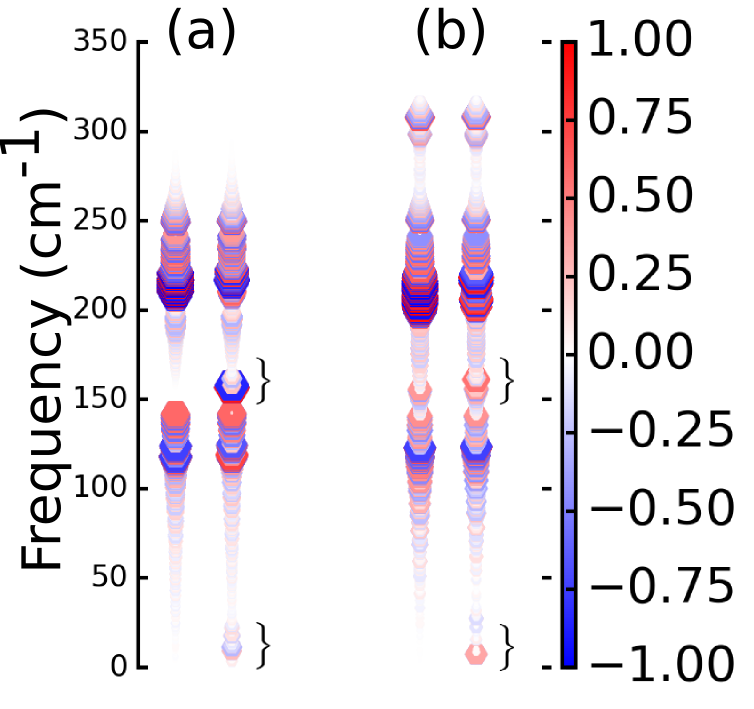}
\caption{(a),(b)-Chiralities at K point for all the phonon modes in 3.3$\%$ biaxially strained moir\'{e} of $\mathrm{WSe_{2}}$ and the $\mathrm{MoSe_{2}/WSe_{2}}$ heterostructure. The left column of (a) denote the non-interacting and the right column of (a) interacting bilayer calculations. Similarly, we show the results for (b). The emergent chiral modes are highlighted by curly brackets. }  
\end{figure}

We apply a bi-axial tensile strain to the bottom layer of an AA stacked bilayer $\mathrm{WSe_{2}}$. The resulting strained structure contains three unique high-symmetry stackings: $\mathrm{A\widetilde{A}}$, $\mathrm{B^{W/\widetilde{Se}}}$ and $\mathrm{B^{Se/\widetilde{W}}}$, where the tilde indicates
a sublattice of the strained layer. These stackings are distinct from those of the twisted bilayer near $0^\circ$ as they break layer symmetry. 
\clearpage
\newpage

\section{VI: Achiral phonon modes at $K/K^\prime$ for twist angles close to $0^\circ$}

\begin{figure}[!ht]
\centering
\includegraphics[width=\textwidth]{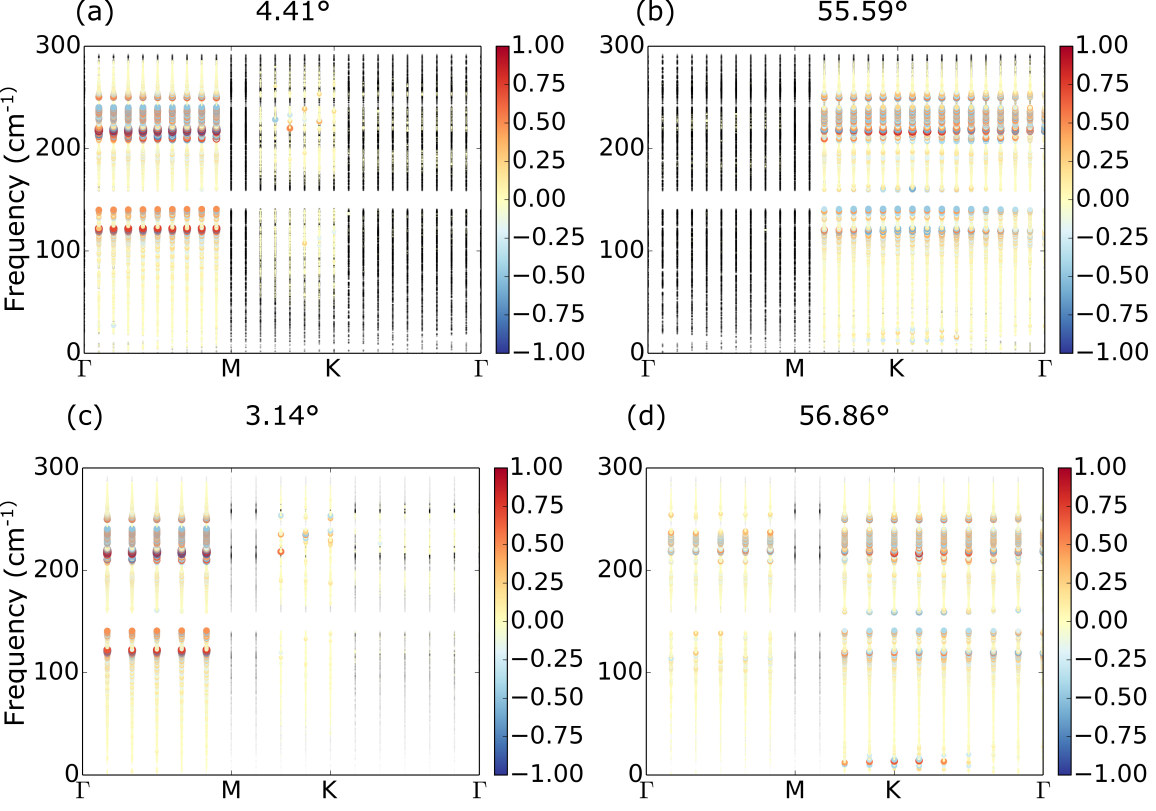}
\caption{Chiral phonons in twisted bilayer $\mathrm{WSe_{2}}$ at twist angle of (a) $4.41^\circ$, (b) $55.59^\circ$, (c) $3.14^\circ$ and (d) $56.86^{\circ}$). Black solid circles represent phonon bands (computed at a few discrete points) and blue and red color indicates the associated chirality (computed at the same set of discrete points). Moreover, the size of the colored circles are proportional to absolute value of the chirality. As this figure is somewhat difficult to interpret, we have also directly inspected the computed chirality values at $\mathrm{K_M}$ and find that there are non-degenerate chiral modes for twist angles close to 60$^\circ$ but none for twist abgles close to 0$^\circ$.}
\label{figS3}
\end{figure}
\clearpage
\newpage

\section{VII: Chirality resolved and total density of states}

\begin{figure*}[!ht]
\centering
\includegraphics[width=\textwidth]{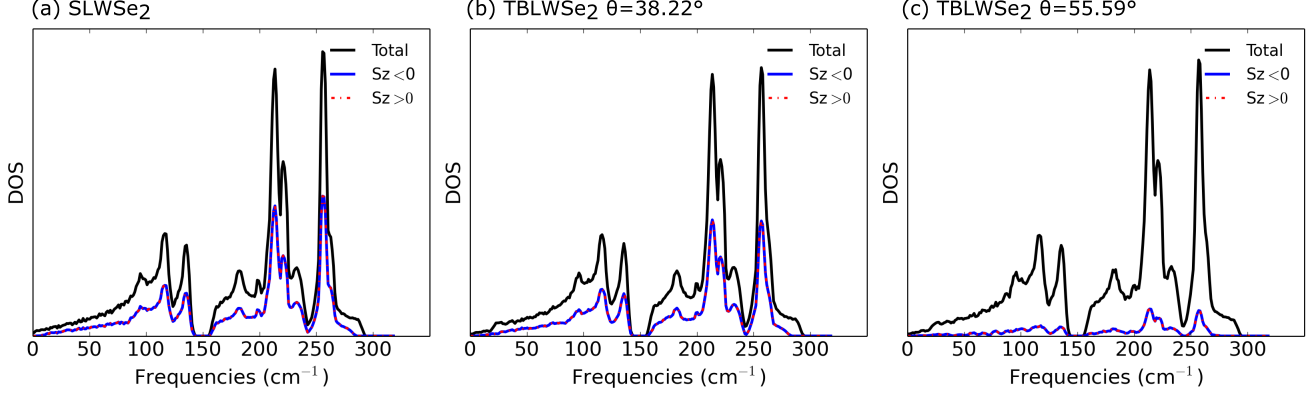}
\caption{(a),(b),(c): Total and chirality resolved density of states (DOS). We have used $36\times36\times1$, $12\times12\times1$, and $3\times3\times1$ grids to compute the phonon DOS for monolayer $\mathrm{WSe_{2}}$, $38.22^\circ$ twisted bilayer $\mathrm{WSe_{2}}$, and $55.59^\circ$ twisted bilayer $\mathrm{WSe_{2}}$, respectively. The delta function in DOS calculation has been replaced by a Gaussian with a standard deviation of 4.8 cm$^{-1}$.}  
\label{figr1}
\end{figure*}

\begin{figure*}[!ht]
\centering
\includegraphics[width=0.8\textwidth]{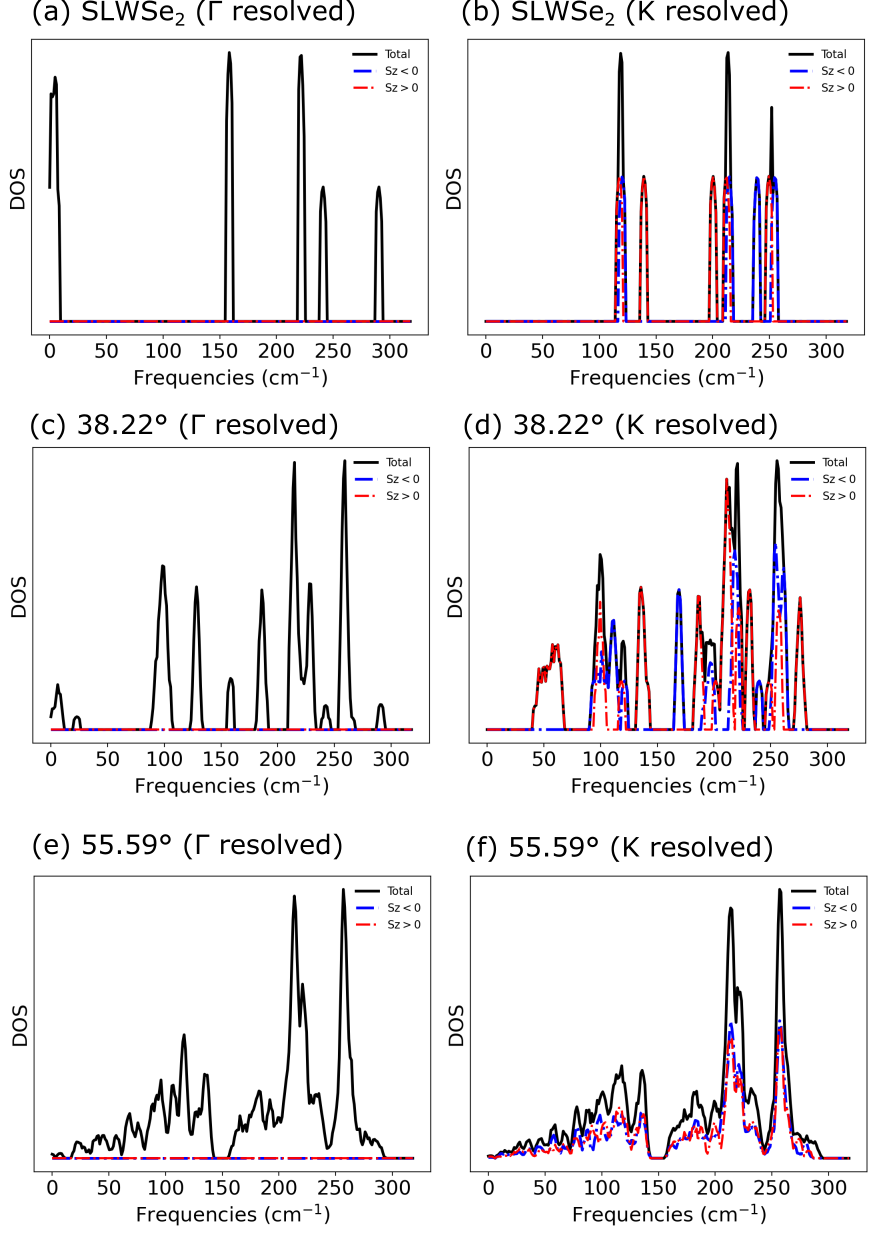}
\caption{(a),(b),(c): $k$-point resolved density of states (DOS). We have used $36\times36\times1$, $12\times12\times1$, and $3\times3\times1$ k-point grids to compute the phonon DOS for monolayer $\mathrm{WSe_{2}}$, $38.22^\circ$ twisted bilayer $\mathrm{WSe_{2}}$, and $55.59^\circ$ twisted bilayer $\mathrm{WSe_{2}}$, respectively. A Gaussian broadening with a standard deviation of 4.8 cm$^{-1}$ has been used.}  
\label{figr1_2}
\end{figure*}
\clearpage
\newpage 

%

\end{document}